# A methodology for co-constructing an interdisciplinary model: from model to survey, from survey to model

Elise Beck[1,5], Julie Dugdale[2,5], Carole Adam[2,5], Christelle Gaïdatzis[3,5], Julius Bañgate[1,2,4,5]

[1]University Grenoble Alpes, Pacte Laboratory, France.
[2]University Grenoble Alpes, Grenoble Informatics Laboratory, France.
[3]Comet', Grenoble, France.
[4]University Le Havre Normandie, LITIS-R2IC, France.
[5]Cogitamus
[1, 2]Elise.Beck │ Julie.Dugdale │ Carole.Adam@univ-grenoble-alpes.fr,
[3]christelle.gaidatzis@gmail.com
[4]julius.bangate@univ-lehavre.fr

## Abstract

How should computer science and social science collaborate to build a common model? How should they proceed to gather data that is really useful to the modelling? How can they design a survey that is tailored to the target model? This paper aims to answer those crucial questions in the framework of a multidisciplinary research project. This research addresses the issue of co-constructing a model when several disciplines are involved, and is applied to modelling human behaviour immediately after an earthquake. The main contribution of the work is to propose a tool dedicated to multidisciplinary dialogue. It also proposes a reflexive analysis of the enriching intellectual process carried out by the different disciplines involved. Finally, from working with an anthropologist, a complementary view of the multidisciplinary process is given.

## 1. Introduction

A growing number of natural disasters (wildfires, storms, earthquakes, tsunamis, etc.) and other crisis events (terrorist attacks, industrial accidents, etc.) occur every year (EM-DAT, n.d.). Society needs to prepare and appropriately manage these events. Therefore, it is essential to understand how the population and individuals are likely to behave in such unpredictable events. Computer science can provide models of human behaviour and simulators to be used as decision support systems to improve crisis management. However, in order to be realistic, these models need to be informed by data, often gathered from social science researchers. Indeed, previous studies have shown the importance of taking into account human behaviour when planning evacuation and population protection after a natural disaster such as an earthquake (Bernardini et al. 2014; Rojo et al. 2017). Sociology also provides insight about the factors of human behaviour that are relevant in disasters. For instance, Mawson's theory of social attachment states that individuals facing a threat will first seek proximity with their attachment figures (family, friends, familiar places and objects, etc.) rather than seeking protection (Mawson 2005). As a result, factors such as social attachment must be taken into account when modelling human behaviour in disasters (anonymised for review).

In previous work, we designed a model of human mobility after an earthquake (anonymised for review). This was a highly multidisciplinary project involving geographers, computer scientists, anthropologists, and field experts. The model was built based on various types of data (both qualitative and quantitative) from various sources, such as official reports, scientific publications, census data and field surveys, etc. Data was collected on: human activities and mobility during earthquakes, especially just after the main shake; social bonds between members of a social group such as between family, friends, and colleagues, according to different European countries; socio-demographic characteristics of the population; population distribution at night and during the day; etc.

Collecting, exploiting and integrating such diverse and multidisciplinary data into a model was an extremely challenging task, especially since no methodology existed at the start of the project to support the process. Therefore, we had to create our own methodology as the project unfolded. The goal of this paper is to share this methodology to help future multidisciplinary projects.

Multidisciplinary research is notoriously difficult and the problems mentioned above are typical of such work. Although integrating empirical data is essential (Boero 2005, §2.17) no standard methodology exists yet and so each project tends to adopt their own. Most often, tasks are split chronologically and by disciplines, leaving the data collection to the social scientists (in our case, social geographers) at the start of the project, while computer scientists deal with model design afterwards. As a result of this task division, there is very little interaction and feedback between data collection and model design. This is far from optimal and can lead to inconsistencies or gaps in the model.

On the contrary, our approach consisted in involving scientists from all disciplines at all stages of the process. We consciously made the decision not to just share information on methods, techniques and procedures used in the various disciplines, but to actively involve all disciplines. This led to more discussions and iterations between the 'data collection' and 'model design' phases, and to an iterative improvement of the model. Concretely, both social scientists and computer scientists participated in the elaboration of a questionnaire for gathering human behaviour data, and both also participated in the design of a conceptual model based on this data. Domain experts then helped with the scenario design and model validation.

In this paper, we describe the multidisciplinary methodology that we devised from our practical experience in this project. We believe that it is generic enough to be applied to and benefit other similar multidisciplinary projects. We focus on explaining how scientists from different disciplines worked together in order to gather data and exploit it to design a more "complete" model. Although we discuss validating the methodology, 'model validation', which ensures that the model is an acceptable representation of reality, is out of the scope of this paper.

The paper is structured as follows. Section 2 surveys the state of the art on model design and data collection. Section 3 describes our approach and the proposed methodology for multidisciplinary data collection via surveys and iterative model design. Section 4 presents the three results obtained from using our methodology: the questionnaire, the model, and the tool to help multidisciplinary dialogue. Finally, we discuss the methodology through a meta-analysis, and conclude this paper in Section 5.

# 2. State of the art

## 2.1 Model design with an interdisciplinary approach

Most efforts to include diverse stakeholders in model design have focused on involving users or domain experts. Indeed, the involvement of stakeholders is now well accepted and commonly used in the design of simulation models. In particular, participatory simulation, in which users are involved very early on in the design of the model, has proved to be particularly advantageous in terms of gaining user acceptance of the final tool and in ensuring that a more complete and refined model is achieved (Ramanath and Gilbert 2004).

One of the first efforts to include different types of disciplinary knowledge in multi-agent systems was Olivier Barreteau's work (Barreteau 1998) involving farmers concerned with the irrigation of the Senegal River. This work developed into coupling multi-agent systems with a role-playing game (RPG). RPGs have been used as a method to elicit knowledge and to formalise assumptions constituting the model. Starting from a basic conceptual model, the RPG is used to improve the formalisation and act as a communication tool between the model and reality (Barreteau 2003; Bousquet et al. 2002).

Companion Modelling (ComMod) is a participatory gaming and simulation approach that uses RPG and simulation models (Bousquet and Trébuil 2005). The ComMod approach is iterative and evolves with the participative process whereby stakeholders are involved in the definition and design of the questions, models, simulations and outputs (Étienne 2013).

RPG and companion modelling predominantly include users or stakeholders that are not designers in order to co-construct and refine the model in an iterative way. Our approach differs in two ways. Firstly, we do not focus on using RPG, but use the survey as an instrument to feed the model. Secondly, although users and domain experts helped in co-constructing the model, we focus on how research scientists from different domains can work together in a complementary way to co-construct the model.

Other works, such as Cioffi-Revilla 2010, use an approach based on conceptualizing and developing a succession of models with increasing complexity as they approximate the target system. This methodology is particularly appropriate for large, multidisciplinary projects for complex social simulations. However, it considers a target to be modelled through different submodels that may be related. In our case, we consider one model being addressed by the different disciplines.

Other works have proposed methodologies to link simulators that stem from different domains (for example agent-based and physics-based) (Kashif et al. 2013). Our work differs in that we are not trying to link simulators, as in a co-simulation approach, but rather to develop one holistic simulator.

## 2.2 Data collection

In order to develop a model of human behaviour it is important, not only to draw upon social theories from the literature, but also to collect data on human activities and characteristics. In our project we were specifically interested in gathering social data on behaviours adopted just after the earthquake, on social bonds, and on the day and night population distribution.

Regarding the latter, spatial and statistical processing of household-travel survey data was extrapolated to give the spatial and temporal distribution of a population over a territory (Roddis and Richardson 1998).

Several methods exist to collect social data on behaviours during natural crises. In our project, gathering social data is highly influenced by the objective of the study. Indeed, among the existing methodologies, some may not be adapted to the thematic framework, i.e. natural unpredictable crises. The literature provides a broad idea of the social response to a natural crisis (Provitolo et al. 2011). We also know that the cultural context plays an important role in risk perception and the behaviours adopted during a natural crisis (Palm 1998; Paradise 2005). Nevertheless, we needed to gather data specifically for the application context of our model.

The collection of behavioural data is possible with: direct observation, indirect observation (video recordings), and interviews or questionnaires.

Direct in situ observation requires a specific protocol covering a social situation and deploying it at a precise moment (Kaplan 2017). The major advantage of direct observation is that the behaviours are not based on declared words (as opposed to surveys by questionnaires or interviews), which makes behaviours more objective. However, the intrinsic unpredictability of earthquakes means that we cannot use this method in this context.

Indirect observation consists in analysing video recordings, provided either by video surveillance or by individuals through digital social networks such as YouTube and FaceBook. This method is powerful to qualitatively describe behaviours (Gu et al. 2013; Lambie et al. 2016; Ziu et al. 2016). However, one of its limitations is the statistical representativity of the output since some extreme behaviours may be over-reported/recorded. Another problem comes from the limited view of the scene; some behaviours may be explained by contextual elements that are not recorded by the camera.

Among the survey techniques, interviews and questionnaires highlight the distinction between qualitative and quantitative approaches. Those two techniques are based on declarations which may bring some subjectivity to the answers.

Surveys via interviews can provide very fine-grained and precise data (Groves et al. 2009). When they are given to small samples, they can uncover in-depth information about a subject. When carrying out interviews, the interviewer seeks to reveal a diversity of opinions or situations. The administration of the survey as well as the processing of answers can be very time-consuming. Among surveys, feedback methods are carried out just after an event. This is done to understand the physical and social causes of the catastrophe and to highlight the dysfunctions in the crisis management in order to improve local prevention methods or national legal framework of risk management. In a scientific context, these methods have been used to collect very precise data on pedestrian mobility behaviours following a flash flood (Ruin et al. 2013) or an earthquake (Rojo et al. 2017), using a spatio-temporal grid or a map (anonymised). Feedback methods, if used to uncover mobility behaviour after an event, may last at least one hour per interviewee.

Like interviews, surveys by questionnaires are widely used to gather social data (De Leeuw et al. 2008), especially behavioural data on medium to large samples. The idea is to obtain a statistically representative picture of a situation at a certain time. Questionnaires can be either administered face-to-face or self-administered, i.e. sent by email, mailed or delivered in mailboxes (Jon et al. 2016). In the case of self-administered questionnaires, the survey-carrier should have verified twice that the questions are well understandable so that the answers correspond to the survey's objective. This can be done by testing the questionnaire several times with various people.

The evolution of ICT and in particular digital social networks means that questionnaires can be easily distributed and completed online. The online accessibility of the survey facilitates its spread through social media, electronic mail, webpages, newsletters, therefore gathering many answers from a wide audience over a very large area. It also allows surveying a more socially diverse sample by reaching people who would not answer a face-to-face survey. Thus, a large number of responses can be easily gathered (Rhodes et al. 2003; Evans and Mathur 2005). For example, more than 1500 answers to a single questionnaire were obtained after the Ml 5.2 Ubaye earthquake in 2014 (Boisson 2015). Since the survey is not addressed to a specific sample, statistical representativity can be pre-assured by using a filter or with statistical adjustment.

The following table summarises the characteristics of the different techniques that can be used to collect social data. Considering modelling needs, online surveys are often the most useful method: they are designed to rapidly gather data over large samples and can be used to collect detailed information if open questions are included.

Table 1: advantages and limitations of different techniques to collect social data.

|  | Objectivity/subjectivity | Statistical representation | Diversity of situations | Time for gathering data | Size of sample | Detailed information |
|---|---|---|---|---|---|---|
| **Indirect observation** | Objectivity | No | No | Long | Medium | Yes but no information on the causes |
| **Surveys by interviews** | Subjectivity | No | Yes | Long | Small | Yes |
| **Online surveys** | Subjectivity | Yes if filtered or re-adjusted | Yes | Fast | Very large | Yes if many open questions |

| Surveys by questionnaire | Subjectivity | Yes | Yes | Long | Medium to large | Yes if many open questions |

## 2.3 Limitations

From the state of the art there are several limitations with current approaches:

- The lack of a tangible tool for use in multi-disciplinary work: a multi-disciplinary approach is increasingly common and necessary, yet few, if any, tools exist to support and facilitate efficient cooperation.
- The difficulty of interactions between computer science and social science: interactions between computer scientists, who develop the simulator, and social scientists, who collect the domain knowledge, are crucial but often difficult. Good tools are needed to help social scientists visualise and understand what is being developed by the computer scientists, and ensure it matches with their domain expertise. Likewise, computer scientists need to understand the intricacies of data collection.
- Limited iterations between data collection and model design: in the absence of such dedicated tools, the iterations will be limited between the scientists of the different fields, often with only a single data collection phase at the start of the project, followed by a single model design phase. If the social scientists are not able to understand the model being designed, they will not be able to comment on it, nor to provide useful feedback to inform its iterative design.
- Existing methods/tools for modelling for computer scientists are often not well understood by social scientists; for example, UML and TDF (Evertsz 2015) have been used for agent modelling (Mancheva et al. 2019; Richiardi et al. 2006) but they are hard to master for non computer scientists.
- Conversely, existing methods/tools for modelling for social scientists are often not well understood by computer scientists; for example, in the social sciences, models can be "designed/described/represented" by paragraphs, tables or graphs obtained from analysis of surveys or even schemes.
- Different vocabulary/tools/methods in different disciplines that have to cooperate. Each discipline has its own vocabulary and some efforts need to be made to facilitate discussions. Experiencing multi-disciplinary cooperation over time helps acquiring new vocabulary from other disciplines or, at least, understanding the way other disciplines define various terms and concepts.

## 2.4 Objectives and proposed approach

To address these gaps in the literature we provide several contributions to the modelling community:
- A methodology, explained in section 3, to bring together approaches from social sciences and computer sciences, for both data collection, via field surveys, and model design.
- An illustration of this methodology, detailed in section 4, for a specific case study of earthquakes. In this context, we designed a questionnaire that is both rigorous from a social science point of view, and useful from a computer science point of view for gathering data required for the model and simulation. We also designed and implemented a simulator based on the survey data and other sources. This model takes into account social interactions in the population after an earthquake[1]).
- A meta-analysis of what can be learnt from applying the methodology is explained in section 4. We believe that the methodology is generic and that it can benefit the modelling community and help them overcome some of the difficulties of multidisciplinary work.

# 3. Method

We first describe our proposed generic methodology in sub-section 3.1. We then analyse how we applied it in a specific project, and discuss the challenges we had to overcome (sub-section 3.2). This meta-analysis aims to provide insights to other researchers wanting to apply our methodology in their projects.

## 3.1 General method - model design

The basic steps of the methodology are shown in Figure 1.

**STEP 1**: The process begins by each team - computer scientists and social scientists - collectively defining the objectives of the model. As noted by Nigel Gilbert this is a crucial step in order to ensure that the developed model directly targets the needs of the stakeholders (Gilbert 2004).

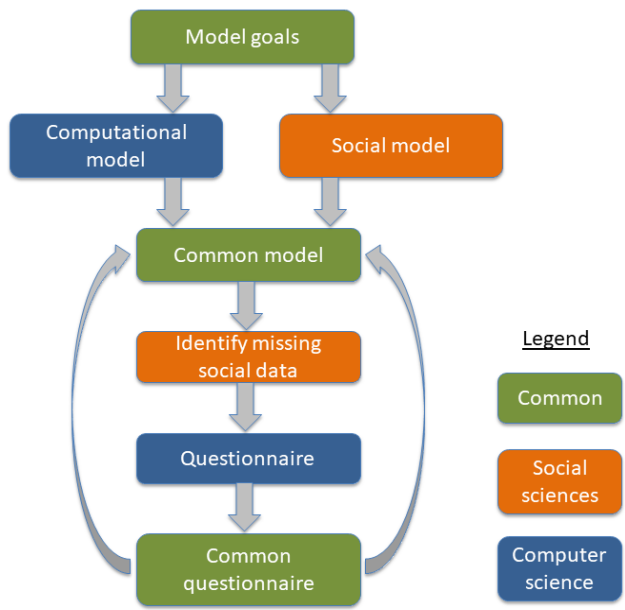

Figure 1: Proposed methodology

**STEP 2**: In the second step, each disciplinary team separately, but in parallel, develops a first draft of the model. In essence, this describes how each team interprets the problem, what critical factors they think should be included, and how these should be conceptualised.

**STEP 3**: Once the two draft models are completed, they are compared in order to identify the common elements of the models, and more importantly the differences: which elements were included in one model but not in the other. This is a highly interactive step with in-depth discussions to explicitly justify the inclusion or exclusion of model elements. The output of this step is a common model, collectively co-constructed by both teams.

---

[1] In the final unanonymised version of this paper, other papers published about this model will be cited.

**STEP 4**: Each element of the model is then examined from the point of view of the social data needed to support those elements. For example if demographic data, such as the age distribution of the population is needed, discussions follow as to how this data will be obtained. This step obviously uncovers cases where elements should be included in the model, but for which no data is yet available to support their inclusion. Obtaining data for model building has been cited as one of the most challenging aspects of constructing a suitable model (Beck et al. 2014). The proposed method allows the teams to clearly identify missing data.

**STEP 5**: Following the identification of the missing data required to feed the model, the computer science team then develops a questionnaire in order to collect this data. The social science team would appear more suited to this task, but there are several arguments behind the choice of having the computer scientists do it. The first is educational, as it exposes the computer science team to the social science culture of data gathering, and abruptly confronts them with the reality of multi-disciplinary work. The second is more practical: questionnaires developed by social scientists typically cover a (too) wide range of research questions, while computer scientists tend to focus on questions that provide data to feed the model. However, if they follow the KIDS approach (Edmonds and Moss 2004) and design a very descriptive model, it may result in an excessively large number of attributes and behaviours being included in the model. The two teams then interact and the social science team's experience is used to revisit the proposed survey, classify research questions from essential to less interesting, remove unnecessary questions to shorten the questionnaire, and rephrase questions when needed. The output of this step is a first version of the questionnaire.

**STEP 6**: The final step of our methodology consists in verifying that the questions in the resulting survey are necessary and sufficient. This step aims at avoiding inconsistencies between model and data that can result from the traditional approach when questionnaire and model are designed separately, i.e. the survey is carried out at the start of the project while the model is designed at the end. Concretely in this step, the teams check that the questionnaire can actually be used to gather all the data that has been identified when conceptualising the model (sufficient questions) and, as an iterative process, that the model has taken into account all the potential behaviours listed in the questionnaire (necessary questions). This validation step checks that the model and survey are consistent with each other, i.e. checking that all the information necessary to the model can be collected by the survey (no relevant question is missing), and that all information collected via the survey is really useful to the model (no irrelevant question is included). In summary, the output of this step is a common questionnaire whose set of questions is both necessary and sufficient to build the model.

## 3.2 Analysis of the multidisciplinary iterative methodology

The process of developing a model using this methodology was analysed by an anthropologist by conducting interviews with the researchers and through direct observation. The goal was to extract the relational dynamics during the process of building a common multidisciplinary methodology. Specifically, we aimed to uncover: 1. How researchers went about jointly reshaping their usual tools, namely the survey questionnaire and the multi-agent model? 2. What were the obstacles to mutual understanding of the disciplines involved and their respective methodological requirements? 3. How these obstacles were overcome or resolved by the team? From the data collected, the longer-term objective was to develop a mediation workshop aimed at fostering interdisciplinary dialogue (see 4.3).

The goal was not so much about observing the content of the information exchanged, but about observing "science being made" (Latour 1989), with its uncertainties and hesitations in a project where reflexivity was key. Indeed, the construction of the multidisciplinary methodology required researchers to analyse their own science, triggered by questions from colleagues from another discipline. Step by step, each scientist explored the methodological tools, ways of categorising the data, and how the problem was interpreted by the other discipline.

Our observation of the lengths and blockages in this process of methodological deconstruction-reconstruction allowed us to identify the sensitive points of interdisciplinary work. Indeed, the research team was already used to working together on the design of multi-agent models of an earthquake crisis. As a result of previous multiple collaborations, the teams had reached a common understanding of some of the tools and concepts of each other's domain (e.g. the notions of "risk perception" and "vulnerability" as used in social geography, and how these may be implemented from the point of view of computer science). This meant that the exchanges were relatively fluid. However, this project involved leaving our comfort zone, in order to closely readjust the methodology to the requirements and constraints of each discipline. The moments of hesitation (in this case the development of a common survey) were informative and highlighted important issues: the critical notion of pre-evacuation delay for computer scientists, or the need for exhaustive identification of individual behaviours after a quake for geographers.

The individual interviews that were conducted with the members of the team compared actual observations of the process with the researchers' feelings about how they learnt, the obstacles encountered, and the methodological choices that were made. They were also asked about their initial motivations for conducting multidisciplinary projects, the usual difficulties of this type of collaboration, and the expected benefits for them. Like other researchers sharing interdisciplinary experience, the team worked together primarily on the basis of intellectual and relational connivance. Multidisciplinary collaboration requires a strong will to overcome disciplinary differences through the development of a common epistemology: the search for unification through a language, a common object, a shared method requires devoting time and many reformulations (Hervé and Rivière 2015). Thus, many of the project exchanges were initiated by "what do you mean by ...?'' questions, aiming at ensuring mutual understanding.

The intrinsic challenges of interdisciplinarity also face another difficulty in constructing and following a project methodology. Beyond science, even though the objective of the work and the steps needed to achieve had been collectively discussed beforehand, they were still potential sources of misunderstanding. In this project, the position of the survey in the methodology and the lack of intermediate steps were questioned *a posteriori* by the researchers. As a social and technical construction process, scientific activity takes the form of collective actions (Aggeri and Hatchuel 2003). The attention paid to mutual understanding at all stages of a project thus goes beyond the question of interdisciplinarity: it aims at understanding more generally the modalities of group interaction.

The work at the heart of a multidisciplinary team finally raises other concerns or even frustrations. These were uncovered during the interviews as: losing one's own identity due to oversimplification, or leaving important concepts aside to facilitate comprehension. The identity of each discipline is at stake here, as well as the limits of interdisciplinary sharing.

Discovering other ways of producing knowledge highlights the division between the disciplines. By breaking down the disciplinary boundaries, it was possible to better understand each of the disciplines, their methodological and conceptual particularities, and the possible bridges between disciplines. As Barth (1969) points out, there is no starting homogeneity, but there is always a relation. Identity is an unstable amalgamation of various self-images derived from relationships with others. If there are multiple challenges in the vision and division of scientific work (Freymond et al. 2003), the identity constructions of the sciences and those who practice them are in perpetual movement, as with any other identity process.

When faced with other disciplines, we make moves towards them in order to facilitate multidisciplinary work. However we also retain a view of our own discipline and this prompts a reflection concerning its specificities, relevance and legitimacy.

There are real distinctions between disciplines, yet the exact divisions are arbitrary, depending on our individual disciplinary perceptions. Because of that, actions of integration, assimilation or acculturation to other disciplines are affected by our view of where those divisions exactly lie. It is the thought of this similarity with intercultural encounters that gave rise to the creation of "The MeetTic of Scientists" workshop described below (see 4.3).

# 4. Results

The multidisciplinary project resulted in a questionnaire, a conceptual model that was implemented as a simulator, and a tool to facilitate multidisciplinary dialogue.

## 4.1 A questionnaire to gather behavioural data

From the limitations and advantages of the ways to collect social data after an earthquake , we set up an online questionnaire. Despite the opportunity to gather a great amount of social data, we focused exclusively on the data that would be useful for modelling individual behaviours in the case of an earthquake. Questions therefore needed to be formulated in such a way that answers could directly be integrated into the model. The structure of the questionnaire allowed people who had not experienced an earthquake to answer the survey. In order to facilitate statistical processing, most of the questions are closed questions.

The questions deal with several aspects:
1) Earthquake experience (if any): the questions aim at relating the past experience of an earthquake, i.e. the spatial and social context of the event, time and intensity of the earthquake, feelings that the event produced, the adopted behaviours (especially those involving mobility) with many details (place where the people went, time to evacuate a building - if relevant, reasons for the evacuation);

2) Preparation strategies and knowledge (people who have never experienced a seismic event can still answer those questions): knowledge of the safe open areas, hierarchy of the social bonds, organisation with kin, neighbours and/or colleagues before or after evacuation, preventative information received from any stakeholder, confidence in the building's resistance to an earthquake;

3) Finally, some socio-demographic questions that characterised the respondent's social profile, for instance age, gender, education, family structure, dependant persons, and mobility capacities.

Regarding genericity, the survey may be used in various geographical and cultural contexts, as it was designed using previous experiences of the authors in France, Italy, Argentina and Lebanon. However, some slight adjustments may be necessary regarding the proposed items for closed questions. The survey is accessible in French, English and Arabic.

As mentioned previously, the online survey is by definition self-administered. For this reason, the questions should be easy to understand by any respondent, which requires testing the questionnaire several times with various people. Despite this disadvantage, the online accessibility of the survey facilitates its spread through social media, electronic mail, webpages, newsletters, therefore gathering many answers from a wide audience over a very large area. It also allows us to survey a more socially diverse sample by reaching people who would not answer a face-to-face survey. Since the survey is not addressed to a specific sample, some statistical adjustments may be necessary if the sample is not representative of the population.

The diffusion of the survey may be achieved by several means. Using the professional and personal networks of the authors may result in an over-representation of the more educated populace. For this reason, posting the URL of the survey on social networks or webpages - especially on local groups (local authorities, municipalities, associations, etc.) - may help diversing the sample.

This survey is accessible online at www.XXX (anonymised for review) and extracts can be found in the annex of this paper.

## 4.2 A model to simulate a seismic crisis at the individual scale

The questionnaire allowed us to gather a lot of qualitative behaviour data, which we then used to design a conceptual model of human social behaviour immediately after an earthquake. The model was built iteratively, with many discussions between researchers from the different disciplines involved. It is now fully implemented, functional, and published (anonymised for review). The focus of this model is on the role of social attachment in individual behaviour just after the main shake of an earthquake. It shows that individuals will first gather with their relatives or friends before evacuating, which can slow down evacuation but also help in getting more people to safety.

## 4.3 A workshop to bootstrap multidisciplinary dialogue

Beyond building this model, we want to answer a more general question. How can the observations and reflections that emerged from this project benefit other researchers from different disciplines? How can they reuse this insight to facilitate their exchanges in other circumstances? Our idea was to create a game or a workshop around this theme in order to facilitate exchanges between scientists on the difficulties encountered during an interdisciplinary cooperation. Without pretending to provide a miracle solution to these difficulties, it did at least open a space for discussion, in order to bootstrap dialogue and mutual understanding that would be required to initiate a research project during its structuration phase.

Multidisciplinary encounters obey the same rules that underlie other forms of human encounters and interactions. This inspired the main thread of the workshop entitled "The MeetTic of Scientists". A first test session was organised over a half-day period in June 2019, and attended by 29 researchers from 10 different domains. The workshop was one of the activities proposed by a summer school dedicated to multi-agent modelling using a spatial approach. However, it could be applied to any other multidisciplinary community wishing to make an introspective analysis of its practices. In the workshop, multidisciplinarity was considered from the point of view of the human encounter, involving the construction of relations between people from different scientific cultures, with all the challenges and benefits that this implies. Combining several techniques from popular education and aiming to encourage the participation of all, the workshop was designed in three stages:

- **Step 1**: the "MeetTIC affinity". In mono-disciplinary groups, participants are asked to share their values, motivations and expectations for interdisciplinary work, and to draw up a short consensual list. Energising and breaking the ice between the participants, this activity seeks to bring out a panel of strong ideas from each group concerning its relationship to interdisciplinarity.
- **Step 2**: "Thematic Speed Dating: Interdisciplinary Relationships and You". This step aims to generate discussion within multidisciplinary teams around defined themes. Teams stay together but move from theme to theme following a predefined schedule (20 minutes per theme). The results of the discussion on each theme are written on posters. The themes deal with multi- or interdisciplinary first experience, setbacks and disappointments, "love at first sight" experiences, discoveries, requirements ("happiness elixir") of each participant for fruitful multi- or interdisciplinary relationships. The idea here is not to generalise or reach a consensus, but to share individual experience. This is followed by an introspective phase, both individual and collective, where individual experiences of interdisciplinary relationships are shared among each group.
- **Step 3**: collective creation of advertisements (Fig. 2 provides some examples) for an interdisciplinary meeting, website or newspaper. If the previous steps aimed at better understanding what made sense for everyone, this more creative phase allows to playfully forge a collective dynamic. In mono-disciplinary groups, participants must find strong arguments to invite their counterparts from other disciplines to break down disciplinary barriers and to engage in mutual collaboration.
- **Step 4**: collective sharing. The different advertisements and posters created during the workshop are displayed to share the experience of each group.

> *If you are a geographer, a biologist, an architect, a sociologist, a computer scientist, or come from any other discipline And you are looking for another point of view (for one night or forever) on your research topic Then we can bring you the power of mathematics that we will co-build Coupling wanted asap*
> *The CS team*

```
Experimented computer
scientists (h-index > 40)
are looking for competent
geographers (if any) for
anthological ontology
project in UML (3.0). Salary
in bitcoins.
```

Are you sick of people discrediting your research? Opposing hard sciences with social sciences? Let's design and validate our COMMON MODEL. Anyway, you won't have a choice...

Figure 2: examples of advertisements for an interdisciplinary meeting website or newspaper.

The workshop is easily replicable and helps to establish a basis for exchange between all participants of a project. It can be of real use in facilitating the start of a new collaboration between researchers from various scientific backgrounds. However, dedicating time to interdisciplinary dialogue is only beneficial if it is renewed multiple times in different forms over the course of the research project. Building a shared vision, and repeatedly adjusting common objectives can only happen over the long term in a reflexive way. The workshop also raised other questions. Interdisciplinarity implies finding a common language to progress together, but what are the real possibilities to share one's ignorance of a subject, a concept, within a group of scientists? To what extent does the search for precision and accuracy, which is undoubtedly one of the strong characteristics of a group of researchers, also result in prohibitively long discussion times for clarifying concepts as required for interdisciplinary dialogue?

# 5. Discussion and conclusion

This original approach produced surprising results. Working in a multi-disciplinary scientific group generated an enhanced attentiveness to each other's work. This reflexive process also provided a step back from our own research practices. As an example, structuring the model was different depending on the discipline: a social scientist arranges the behaviours of the people taking into account their wish of changing their initial activity plan, or depending on whether the people may move or not. Conversely, a computer scientist structures the model with a view to coding/implementation and taking into account different possible architectures to model these wishes.

This methodology meant that the time to develop the model was longer than with the traditional approach and it required a lot of energy from both disciplines. However, it has helped the researchers involved to get accustomed to and to appropriate both the methodologies (model design and formalisation and questionnaire) and the results (answers to the questionnaire and model obtained). They also improved, not only their multidisciplinary approach, but the consistency of the implemented model.

In conclusion, this paper has made the following contributions:
- A novel generic methodology for multidisciplinary model design
- A methodology to design a questionnaire dedicated to model conceptualisation

- A meta-analysis of the multidisciplinary methodology applied to a specific project (earthquakes)
- The design of a workshop to help bootstrapping multidisciplinary dialogue

We obtained results for a specific model of earthquakes in the context of a particular project (anonymised for review), but we also showed that our methodology is beyond this unique project, and generalisable. We intend our methodology to benefit other researchers engaged in similar multidisciplinary projects, by saving them time and removing the obstacles involved in this complex but fulfilling process.

# Acknowledgements

The research was financially supported by ARC7 (Région Rhône-Alpes) and IXXI.

# Annex

Extracts from the questionnaire

## Your experience of earthquakes

**Have you experienced an earthquake? (if several, choose the strongest you can remember)**
- Yes
- No

**Where were you?**
- In my country of residence
- In a familiar country
- In a foreign country

**Where exactly were you?**
- At home
- At work
- With friends / family
- At school / place of study
- In a public building (library, store, swimming pool, etc.)
- In a park, a place or a field
- On the road
- I do not remember
- Other

### How did you feel during the earthquake?
- [ ] I ignored the shake and continued what I was doing
- [ ] I did not have time to react
- [ ] I panicked, I was frozen with fear or I cried
- [ ] I thought about the different options
- [ ] I appreciated, took advantage of the moment
- [ ] I hope there is no other jolt
- [ ] I waited for the shaking to end
- [ ] I prayed
- [ ] Other

### Do you feel like you've done the right thing?
- ○ Complètement
- ○ Partially
- ○ Not at all
- ○ I do not know

### Did you follow alert or evacuation orders?
- ○ Yes
- ○ No

### Do you know where the closest safe area to your home and / or workplace is?
- ○ Yes, I know the safe area near my work
- ○ Yes, I know the safe area near my home
- ○ Yes, I know the safe area near my work and my home
- ○ No I do not know

### How do I get to the safe zone?
- ○ Walk
- ○ A moto/scooter
- ○ By bike, scooter, skate …
- ○ By bus, taxi …
- ○ By car
- ○ Other

### Do you know what to do in the event of an earthquake if you are …

|  | Yes | No |
|---|---|---|
| …at home | ○ | ○ |
| …in the street | ○ | ○ |
| …at work | ○ | ○ |

## Who / what are you most concerned about in the event of an earthquake or comparable natural situation? rank options 1 to 6, with 1 being the most important

|  | 1 | 2 | 3 | 4 | 5 | 6 |
|---|---|---|---|---|---|---|
| Children | ○ | ○ | ○ | ○ | ○ | ○ |
| Husband / wife / partner | ○ | ○ | ○ | ○ | ○ | ○ |
| Parents) | ○ | ○ | ○ | ○ | ○ | ○ |
| Brother sister | ○ | ○ | ○ | ○ | ○ | ○ |
| Neighbors | ○ | ○ | ○ | ○ | ○ | ○ |
| Other family members | ○ | ○ | ○ | ○ | ○ | ○ |
| Friends | ○ | ○ | ○ | ○ | ○ | ○ |
| Colleagues | ○ | ○ | ○ | ○ | ○ | ○ |
| Pet or farm animal | ○ | ○ | ○ | ○ | ○ | ○ |
| Objects, house | ○ | ○ | ○ | ○ | ○ | ○ |
| Unknown | ○ | ○ | ○ | ○ | ○ | ○ |

## Would you wait for everyone in your accommodation to leave the building?

○ Yes all
○ Yes almost all
○ No I would not wait for anyone